\documentclass[12pt]{JHEP}
\usepackage{amsmath,amssymb}
\usepackage{yfonts}

\newcommand{\be}{\begin{equation}}
\newcommand{\ee}{\end{equation}}
\newcommand{\beqa}{\begin{eqnarray}}
\newcommand{\eeqa}{\end{eqnarray}}
\newcommand{\bea}{\begin{eqnarray}}
\newcommand{\eea}{\end{eqnarray}}

\newcommand{\HHH}{{\cal H}}
\newcommand{\w}{{\omega}}

\newcommand{\wn}{\textswab{w}}

\def\d{\partial}

\def\eeq{\end{equation}}

\def\au{ \alpha_1}
\def\ad{\alpha_2}
\def\at{ \alpha_3}

\relax

\preprint{ {\tt hep-th/0601144}}

\title{Shear viscosity from R-charged AdS black holes}

\author{ Javier Mas 
\\

 Departamento de F\'\i sica de Part\'\i culas,
Universidad de Santiago de Compostela \\ E-15782 Santiago
de Compostela, Spain\\
e-mail: jamas@fpaxp1.usc.es
}

\abstract{We compute the shear viscosity in the supersymmetric Yang-Mills  theory dual to the STU background. This is an example of thermal gauge theory with a nonzero chemical potential. The quotient of the shear viscosity over the entropy density  exhibits no deviation from the well known result $\eta/ s = 1/4\pi$. }


\begin{document}

%%%%%%%%%%%%%%%%%%%%%%%%%%%%%%%%%%%%%%%%%%%%%%%%%%%%%%%%%%%%%%%%%
%%%%%   Section 1. Introduction
%%%%%%%%%%%%%%%%%%%%%%%%%%%%%%%%%%%%%%%%%%%%%%%%%%%%%%%%%%%%%%%%%
\section{Introduction}

The AdS-CFT correspondence is a calculational scheme that allows to obtain results in strongly coupled gauge theories \cite{Maldacena:1997re}.
The extension to asymptotically AdS  spacetimes with a regular horizon is relevant 
in connection with the thermodynamical propeties of the dual gauge theory at  finite temperature.
 Equilibrium properties match up to numerical factors \cite{Gubser:1996de}. Near equilibrium, the low energy behaviour should be governed universally by hydrodynamics. A program to obtain transport coefficients was initiated in \cite{Policastro:2002se}\cite{Policastro:2002tn}
 and a number of results have been obtained since then. The upshot of these calculations was a rather peculiar universal behaviour for the ratio of  shear viscosity $\eta$ and  entropy density $s$ of the associated plasma. In all the examples analized, the result
\be
\frac{\eta}{s} = \frac{1}{4\pi} \label{etas}
\ee
was found, and in \cite{Kovtun:2004de} this persistence was related to the universality of the low energy absortion cross section of gravitons  \cite{Das:1996we}. In  the  context of the AdS-CFT formalism, a proof involving the holographic evaluation of correlators of the energy-momentum tensor was presented in \cite{Buchel:2004qq}. It extended the above result to supergravity backgrounds for which the relation $R^t{_t} = R^{x_i}{_{x_i}}$ (no sum) among components of the Ricci tensor holds. A significant class of exceptions to this condition include backgrounds which are dual to ${\cal N}=4$  $SU(N)$ supersymmetric Yang-Mills at finite temperature and with a nonzero chemical potential for the $U(1)^3\subset SO(6)_R$ R-charge. The field strengths of the abelian gauge fields support the difference among components of the Ricci tensor
$R^t{_t} -  R^{x_i}{_{x_i}}\sim F^I_{rt}F^{I\,rt}$.
One such example is the so called STU model, a solution of five dimensional  $N=2$, $U(1)^3$  gauged supergravity first found in \cite{Behrndt:1998jd}. Indeed,  in \cite{Cvetic:1999xp} a particular case of this solution was seen to be obtainable from a consistent Kaluza Klein reduction of $D=10$ type $IIB$ supergravity of  a stack of  black branes that rotate in the internal $S_5$, in the near horizon approximation.

The aim of this paper is to computationally fill this  small  gap.
Calculations have been  performed by means of the Kubo relation, which yields a direct
expression of the shear viscosity in terms of  retarded correlators of the energy momentum tensor. For the quotient  $\eta/s$ we find the result (\ref{etas}) is shown to persist, a fact which supports the extension  
to more general backgrounds than those considered in \cite{Buchel:2004qq}.

%%%%%%%%%%%%%%%%%%%%%%%%%%%%%%%%%%%%%%%%%%%%%%%%%%%%%%%%%%%%%%%%%
%%%%%   Section 2. The STU black hole
%%%%%%%%%%%%%%%%%%%%%%%%%%%%%%%%%%%%%%%%%%%%%%%%%%%%%%%%%%%%%%%%%
\section{Thermodynamics of the  STU background}
In this section we shortly review the thermodynamics of the STU model and set up the conventions.
Let us start by writing the bare action of five dimensional $N=2,  U(1)^3$ 
gauged supergravity.
 
\bea
I_0=\frac{1}{2\kappa^2} \int d^5 x  \sqrt{g}\left( -R   - \frac{4}{L^2} \sum_{I=1}^{3}  e^{\vec a_I\cdot\vec \phi} +\frac{1}{2}(\partial \vec\phi)^2 
  + \frac{1}{4}\sum_{I=1}^{3} 
e^{2\vec a_I\cdot\vec \phi} (F^{I})^2  
-  \frac{\epsilon^{\mu\nu\rho\sigma\lambda}}{4\sqrt{g}}    F^1_{\mu\nu}F^2_{\rho\sigma} A^3_\lambda \right)
   \label{bareact}
\eea
where $\vec \phi = (\phi_1,\phi_2)$,  $\vec a_1 = (\frac{1}{\sqrt{6}} ,\frac{1}{\sqrt{2}}), \vec a_2=(\frac{1}{\sqrt{6}} ,-\frac{1}{\sqrt{2}})$
and $\vec a_3 = (-\frac{2}{\sqrt{6}},0)$.
The  STU solution depends on two functions of four parameters $\mu$ and $q_I~ (I=1,2,3)$.
\be
H_I(r) = \left(1 + \frac{q_I}{r^2}\right) ~~~~~;~~~~~\HHH(r) = \prod_{I=1}^3 H_I (r)
~~~~~~;~~~~~~ f(r) =   k -\frac{\mu}{r^2} +  \frac{r^2}{L^2} \HHH(r)
\ee
with which we can write all the field dependences. For example, the metric tensor assumes the form 
\be
ds^2_5 =-\HHH(r)
^{-2/3} f(r) dt^2 + \HHH^{1/3}(r)\left( f^{-1}(r) dr^2 + \frac{r^2}{L^2} \,d\Sigma_{3,k}^2 \right) 
\label{STUr}
\ee
and  the scalar and gauge fields exhibit the following profiles
\be 
 \phi_1= \frac{1}{\sqrt 6} \log H_1 H_2 H_3^{-2} ~~~;~~~
\phi_2 = \frac{1}{\sqrt 2} \log H_1 /H_2 ~~~;~~A^I_t \,=\, \sqrt{\frac{\mu  }{q_I} + k}~
\left(1-H^{-1}_I\right)
\ee
The discrete parameter $k=0,\pm 1$  controls the choice of the spatial slices $\Sigma_{k}$ of 
constant curvature
\bea
d\Sigma_{3,k}^2 \equiv \eta^{(k)}_{ij} dx^i dx^j = \left\{
\begin{array}{ll}
L^2 (d\theta_1 + \sin^2\theta_1 \,d\theta_2 + \sin^2\theta_1\sin^2\theta_2 \,d\theta_3) & ~~\hbox{for}~k=+1 \\
dx^2+ dy^2 + dz^2 & ~~\hbox{for}~k=0 \\
%:
L^2 (d\theta_1 + \sinh^2\theta_1 \,d\theta_2 + \sinh^2\theta_1\sin^2\theta_2\, d\theta_3) & ~~\hbox{for}~k=-1 \\
\end{array}
\right.
\eea
The case $k=0$  can be uplifted to the near horizon metric for a stack of plane parallel branes that rotate in the internal $S^5$ with angular momenta
 proportional to the charges \cite{Cvetic:1999ne}\cite{Cvetic:1999xp}. In the following section we shall investigate this case, but
 for completeness, in this section we  give  expressions that encompass the three situations $k = 0,\pm1$, (to our knowledge these have  not appeared elsewhere).
We shall denote the volume of the space $d\Sigma_{3,k}^2$ as
\bea
V_{3,k }
= \left\{
\begin{array}{ll}
 2\pi^2 L^3 & ~~\hbox{for}~k=+1 \\
\int d^3x &~~\hbox{for}~k=0 \\
4\pi L^3 \int \sinh^2\theta d\theta &~~\hbox{for}~k=-1 \\
\end{array}
\right.
\eea
  It will be convenient to trade the nonextremality parameter  $\mu$ for the horizon radius, $r=r_+$ given as the largest root of  $f(r_+)=0$  or
\be
\mu = r_+^2 \left(\frac{r_+^2}{L^2 } \HHH (r_+)   + k\right)\, .
\ee
The entropy  density $s = S/V_{3,k}$ is given by the area of the horizon
\be
s = \frac{2\pi }{\kappa^2} A= \frac{2\pi  }{\kappa^2 \,L^3}\sqrt{\prod_{I=1}^3(r_+^2 + q_I)}~,
\label{entropy}
\ee
and for the Hawking temperature one finds
\be
T =  
\frac{1}{2\pi L^2}
   \frac{2r_+^6 + (k L^2 +\sum_{I=1}^3 q_I)r_+^4-\prod_{I=1}^3q_I}{r_+^2\sqrt{\prod_{ I=1}^3 (r_+^2 + q_I)}}.
\ee
There is also  a chemical potential conjugate to the physical charge 
\be
\tilde q_I^2 =   q_I(r_+^2 + q_I)\left(\frac{1}{L^2 r_+^2}
 \prod_{J\neq I} (r_+^2+q_J)+ k \rule{0mm}{3.5mm}\right) 
\ee
given by the gauge field evaluated at the horizon
\be
\Phi^I = \left.\frac{1}{\kappa^2} A^I_t(r)\right\vert_{r=r_+} = \frac{1}{\kappa^2 } 
\frac{\tilde q_I}{  r_+^2 + q_I}
\ee

The thermodynamics of the STU black hole solution has been examined in depth
in the past   \cite{Behrndt:1998jd}\cite{Cvetic:1999ne} where conventional substraction schemes were used in order to extract finite quantities from the asymptotically AdS metric.
In \cite{Buchel:2003re}\cite{Liu:2004it}\cite{Caldarelli:2004ig} the subject was revised from the point of view of the holographic AdS-CFT renormalization prescription.  
The holographic renormalization of asymptotically AdS spaces 
 is by now fairly well understood (see \cite{Skenderis:2002wp} and references therein). The addition of a set of covariant boundary counterterms render the action and the correlation functions finite. A nice feature is that these
  only depend upon the theory under consideration and not the particular solution one
  is interested in.  
 For  pure gravity the set of necessary counterterms has been classified in dimensions up to $d+1=7$ \cite{Emparan:1999pm}. In the present situation there is  a bunch of additional fields present. A systematic construction for an action 
 like the  one here was accomplished in 
\cite{Batrachenko:2004fd} (whose conventions we follow) using the Hamilton-Jacobi method of \cite{deBoer:1999xf} . The result can be written as
\be
I = I_{0}+I_{GH} + I_{ct} \label{renact}
\ee
where
\bea
I_{GH} &=&\frac{1}{\kappa^2} \int_{\Sigma_u} d^4 x \sqrt{-h}\, K  \nonumber\\
I_{ct} &=& \frac{1}{\kappa^2} \int_{\Sigma_u} d^4 x \sqrt{-h} \left( W(\phi) + \frac{L}{4} R + {\cal O}(R^2)\right) \label{ctterm}
\eea
with $K$ the trace of the extrinsic curvature and $h_{\mu\nu}$ and $R$ the 
induced metric and Ricci scalar on the boundary.
 $W(\phi)$ is  the superpotential satisfying
\be
V = 2 \sum_{i}\left(\frac{\d W}{\d \phi^i}\right)^2-\frac{4}{3}W^2.
\ee
This result also appears in \cite{Papadimitriou:2004ap} a similar  setup. In the present case
\be
W(\phi) =   \frac{1}{L}\sum_{I=1}^3 e^{-\vec a_I \vec \phi}.
\ee
We start by listing here the relevant results for the STU background. For the renormalized action  
\be
I_{ren} = \frac{V_{3,k}}{2\kappa^2 L^2 T} 
\left( k r_+^2 + \frac{3}{4}k^2  L^2 - \frac{1}{L^2r_+^2}\prod_{I=1}^3
(r_+^2 + q_I)\right) 
\ee
and for the energy momentum tensor 
\bea
T_{tt}& =&    \frac{1 }{2\kappa^2  L } \left( \frac{12}{L^2r_+^2} \prod_{I=1}^3
 (r_+^2 + q_I) + k \left(12 r_+^2 + 8\sum_{I=1}^3 q_I \right)+ 3 k^2 L^2 	\right)\frac{1}{r^2}
\nonumber \\
T_{ij}  &=& \frac{1}{3} T_{tt}\, \eta^{(k)}_{ij}  .\nonumber 
\eea
With this, we can easily obtain the energy density  
$$
\epsilon =  \frac{1 }{8 \kappa^2    L^3} \left( \frac{12}{L^2r_+^2}\prod_{I=1}^3
 (r_+^2 + q_I) + k \left(12 r_+^2 + 8\sum_{I=1}^3 q_I   \right)+ 3 k^2 L^2 	\right).
$$
Making $T I_{ren} =  gV_{3,k}$ identifies $g$ with the Gibbs free energy density of the associated grand canonical ensemble, and one can easily check that the expected thermodynamic relations hold 
\bea
g  &=& \epsilon - T s - \sum_{I=1}^3\tilde q_I \Phi^I \nonumber\\
 dg &=& -s dT -\sum_{I=1}^3\tilde q_I d \Phi^I 
\eea

\section{Shear viscosity from scalar perturbations}

In this section we shall set $k=0$.  
 The above considerations allow us to derive readily some results for transport properties from equilibrium data. We observe that the energy momentum tensor is traceless, hence with $P = T_{ii}$ the equality
 $\epsilon  =  3P$ leads to the conformal value for the speed of sound 
 \be
 v_s =\left( \frac{\partial P}{\partial \epsilon} \right)^{\frac{1}{2}} = \frac{1}{\sqrt 3}
 \ee
as well as to vanishing bulk viscosity $\zeta= 0$ \cite{Landau}. 
Let us turn to  a new radial coordinate $u = (r_+/r)^2$ and, after  defining   
$a_I = q_I/r_+^2$,  the STU background becomes
\be
ds^2_5 =-\HHH(u)
^{-2/3} f(u) dt^2 + \HHH^{1/3}(u)\left( f^{-1}(u) \frac{r_+^2}{4 u^3} du^2 + \frac{r_+^2}{u L^2} \,d\vec x^2
\right)  \label{STUu}
\ee
\bea
\HHH(u)& \equiv &  \prod_{I=1}^3 (1+ a_I u) \equiv 
1 + \au u + \ad u^2 + \at u^3 \nonumber
\\
f(u) &=& \frac{r_+^2}{u L^2}\left(\HHH(u) -u^2 \HHH(1) \right) \nonumber
\eea
In field theory there are several strategies to compute the shear viscosity. 
Probably the  most straightforward one is to make use of  
Kubo's relation 
\be
\eta = \lim_{\w\to 0}\frac{1}{2\w i }
  \left( G_{xy,xy}^A(\w,0)  - G_{xy,xy}^R(\w,0)  \right)  \label{Kubo}
\ee
where the retarded Green's function is given by
\be
G^R_{\mu\nu,\lambda\rho}(k) = -i \int d^4 x e^{-i k\cdot x} \theta(t)
\langle [ T_{\mu\nu}(x),  T_{\lambda\rho}(0)]\rangle
\ee
and $G^A_{\mu\nu,\lambda\rho}(k)  = G^R_{\mu\nu,\lambda\rho}(k)^* $.
Whereas the original AdS-CFT was designed for Euclidean AdS bulk metrics, the computation of retarded Green«s functions only makes sense in Minkowskian AdS.
A heuristic prescription to compute the retarded two-point function was put forward in \cite{Son:2002sd}.
In order to make use of the Kubo formula, we have to set up a perturbation of the form 
$h_{xy}(u,x^\mu)$ to the metric and compute the on-shell action as a functional of its boundary value $h_{xy}(0,x^\mu)$. It  is easy to check on the equation of motion that the variation of all other supergravity fields can be consistently set to zero.
A convenient parametrization is given in terms of $\varphi(u,x^\mu) = g^{xy}  h_{xy}  $.
 At linearized order the equation of motion for this polarization is nothing  but the
 equation for a minimally coupled scalar. In a Fourier basis
 $\varphi(u;x^\mu) = e^{-i \w t } \varphi_\w (u)$ we find
  \be
 \varphi''_\w + \frac{1+(1+\au+\at) u^2-2\at u^3}{u(u-1)(1+(1+\au)u-\at u^2)}\varphi'_\w + 
 \frac{\HHH(u) \,\wn^2}{u(u-1)^2  (1+(1+\au)u-(\at)^2 u^2)^2}\varphi_\w  = 0.
 \ee
where  $\wn = \frac{\w L^2}{2 r_+}$.
Given the asymptotic normalization  $ \varphi_\w (0)=1$, a regular solution to this equation that satisties the incoming boundary conditions at the horizon $u=1$ can be found perturbatively in $\wn$
 \bea
 \varphi_\w (u) &=& (1-u)^{- i \wn \Gamma}
 \left[ \rule{0mm}{6mm} 1+ \right.
 \\  \label{regsol}
 &&  \left.
 +\frac{i}{2}  \wn\Gamma 
 \left(\Delta  \log 
  \frac{(\Xi -\au -1 +2\at u)}{(\Xi +\au +1 -2\at u)}
 \frac{(\Xi +\au +1)}{(\Xi -\au -1)} + \log
 (1 + (\au+1)u -\at u^2) \right)
  + {\cal O}(\wn^2) \right]
\nonumber
 \eea
where the following definitions have been used
$$
\Gamma =  \frac{\sqrt{1+\au+\ad+\at}}{2+\au-\at}~~~;~~~
\Xi = \sqrt{ 1+\au(2+\au)+4\at} ~~~;~~~
\Delta = -\frac{3+\au}{\Xi } ~.
$$

Expanding the right hand side of the renormalized action (\ref{renact}) up  to second order in $\varphi(x,u)$
and expressing the result in terms of the Fourier transform
\be
\varphi(x,u)  = \int \frac{d^4k}{(2\pi)^4} e^{ i kx}  f(k) \varphi_k(u)
\ee

we obtain the following contributions at the regulating surface $\Sigma_u$
\bea
I_0 &=& \frac{1}{2\kappa^2} \int \frac{d^4k}{(2\pi)^4}f(k) f(-k)  \int_0^1 du (A \varphi''_k \varphi_{-k} + B \varphi'_k\varphi'_{-k} + C\varphi'_k\varphi_{-k} +D \varphi_k\varphi_{-k} )
\label{icero}  \\
I_{GH}  &=& \frac{1}{2\kappa^2} \int \frac{d^4k}{(2\pi)^4}f(k) f(-k) 
\left( H \varphi_k \varphi_{-k} + I \varphi_k' \varphi_{-k} 
\right) \label{igh}
\label{igh} \\
I_{ct}  &=& \frac{1}{2\kappa^2} \int \frac{d^4k}{(2\pi)^4}f(k) f(-k) ~
  J \varphi_k \varphi_{-k}  \label{iW}
\label{ivol} 
\eea
where the functions $A,B,C,D,H,I,$ and $J$ are given in the appendix. Once the perturbations are  set on shell, the bulk action  becomes a   surface term \cite{Buchel:2004di}:
\be
I_0 =\frac{1}{2\kappa^2} \int \frac{d^4k}{(2\pi)^4}f(k) f(-k) \left(  \int_0^1 du [E.O.M]\varphi_{-k}+ 
\left. \left( \frac{C-A'}{2} \varphi_k\varphi_{-k} + B\varphi'_k\varphi_{-k}\right)\right\vert^1_u\right) \label{i0}
\ee
hence all  contributions be arranged in the
form of  pure boundary terms 
$$
I_A = \int \frac{d^4k}{(2\pi)^4} f(k) f(-k) \left. {\cal F}_A(k,u)\right\vert^1_0
$$
The  prescription for computing retarded Green«s function as given in 
\cite{Son:2002sd} is
\be
G^R_{xy,xy}(k) =  -2 {\cal F}(k, u=0)
\ee
where $ {\cal F}(k, u) = \sum_A  {\cal F}_A(k, u)$.
Inserting the solution (\ref{regsol}) into  (\ref{igh}) (\ref{iW}) and  (\ref{i0}), we obtain
\bea
{\cal F}_0(k, u) &=&  \frac{r_+^4}{2\kappa^2 L^5} \left[ \frac{1}{u^2} + 
\frac{2\au}{3u}  -   \frac{3(1+\au+\at)+2\ad }{3}- 3i\sqrt{1+\au+\ad+\at}\,\wn   
 +  {\cal O}(\wn^2,u) \right]\nonumber \\
{\cal F}_{GH}(k, u) &=& \frac{r_+^4}{2\kappa^2 L^5} \left[-\frac{4}{u^2} - \frac{8\au}{3u}  + \frac{6(1+\au+\at)+2\ad }{3} +4 i \sqrt{1+\au+\ad+\at}\, \wn   +  {\cal O}(\wn^2,u) \right] \nonumber \\
{\cal F}_{ct}(k, u) &=& \frac{1}{2\kappa^2} \frac{r_+^4}{L^5} \left( \frac{3}{u^2} + \frac{2\au}{u} -\frac{3(1+\au+\at)+\ad}{2}      +  {\cal O}(\wn^2,u) \right) \nonumber
\eea
 Adding up we   see that the solution is properly renormalized and  finite when $u\to 0$  as expected \footnote{the contribution of the second counterterm in 
(\ref{ctterm}) starts  at ${\cal O}(\wn^2)$}. Moreover we get the retarded
 Green's function to that order
 \be
 G^R_{xy,xy}(\w) =  \frac{1}{2\kappa^2} \frac{r_+^4}{L^5}\left((1+\au+\ad+\at)
 -2 i\sqrt{1+\au+\ad+\at}\,\wn +  {\cal O}(\wn^2)
 \rule{0mm}{4mm}\right) 
 \ee
  Inserting this expression into the Kubo relation (\ref{Kubo}) gives the 
 the following result for  the shear viscosity   \be
 \eta = \frac{1}{2\kappa^2 L^3} \sqrt{\prod_{I=1}^{3}(r_+^2+ q_I)}
 \ee
One  may wish to translate this   into QFT language by uplifting to $IIB$ supergravity and using the
  standard dictionary
  \be
\frac{1}{2\kappa^2} = \frac{V_5}{2\kappa_{10}^2}  =
\frac{ N^2}{8\pi^3 L^3}
\ee
In view of (\ref{entropy}) we also recover the result (\ref{etas}), as promised.

\section{Conclusion}

We see that the proposed holographic viscosity bound \cite{Kovtun:2004de}
is also saturated  in supergravity backgrounds whose dual CFT have a nonvanishing
chemical potential.

Just as an aside, 
  in \cite{Kovtun:2003wp} a closed expression for the shear viscosity was proposed relying
on the so  called ``membrane paradigm". 
Although not rigorously obtained from first principles, this expression is nice both for its simplicity and because
it involves properties of the metric close to the horizon. 
In \cite{Buchel:2003tz} it was shown  that this closed formula reproduced
the  universal result  (\ref{etas}) when restricted again to the class of supergravity backgrounds for which
$R^t{_t} = R^{x_i}{_{x_i}}$. 
For the STU geometry, taken plainly, this expression  apparently signals a deviation of the  above mentioned quotient $\eta/s=1/4\pi(1+...)$. However this formula is not applicable in the present
context, as it is based upon the assumption that metric perturbations in the shear channel decouple, whereas in the STU background  they do couple to the gauge field. It would be nice to find a modification  of that formula that could encompass such mixing.

 While this work was in progress we were informed by  A. Starinets  about a project
  which overlaps significantly with the one presented here \cite{Son:2006em}. Also
  O. Saremi  has worked out the shear viscosity in the presence of chemical potential in the context of M-theory backgrounds \cite{Saremi:2006ep} (see also
  \cite{Maeda:2006by}).

 \section*{Acknowledgments}
I would like to express my gratitude to Andrei Starinets for sharing with me his insight on this topic. Also want to thank   Carlos Nu\~nez for  drawing my attention to the STU background, and  to Roberto Emparan,  Kostas Sfetsos and  Kostas Skenderis for comments. 
The present work has been supported by MCyT, FEDER and Xunta de Galicia under grant FPA2005-00188, and
by EC Commission under grants HPRN-CT-2002-00325 and MRTN-CT-2004-005104.

\appendix

%%%%%%%%%%%%%%%%%%%%%%%%%%%%%%%%%%%%%%%%%%%%%%%%%%%%%%%%%%%%%%%%%
%%%%%   Appendix A. General Solution for Intersecting Black-Branes
%%%%%%%%%%%%%%%%%%%%%%%%%%%%%%%%%%%%%%%%%%%%%%%%%%%%%%%%%%%%%%%%%
\section{Coefficients of the renormalized action}

The coefficients that enter the renormalized action (\ref{icero}) (\ref{igh}) and (\ref{ivol})
are given by the following expressions

\bea
A &=& -\frac{r_+^4}{L^5 u} \left[4(1-u)(1+(\au+1)u -\at u^2)\right] \nonumber \\
B &=& -\frac{r_+^4}{L^5 u} \left[3(1-u)(1+(\au+1)u -\at u^2)\right]\nonumber \\
C&=& \frac{r_+^4}{ L^5}\frac{1 }{3 u^2 \HHH(u)}
\left(\rule{0mm}{6mm}-24 \at^2 u^6 + 2\at(6(1+\au+\at)-11\ad)u^5 + 
10((1+\au+\at)\ad-2\au\at)u^4  \right. \nonumber\\
&& \left. \rule{0mm}{6mm}~~+ (8\au(1+\au+\at)+2\au\ad-6\at)u^3 +
(4\au^2 + 6(1+\au+\at)+14\ad)u^2 + 22\au \, u + 18\right) \nonumber \\
D &=& -\frac{r_+^4}{L^5}\left[
\frac{\HHH(u)\wn^2}{u^2(1-u)(1+(\au+1)u-\at u^2)} +
\frac{1}{3u^3 \HHH(u)^2}\left( \rule{0mm}{6mm}
(\at\ad(1+\au+\ad+\at)-2\at^2\au)u^7
\right.\right.
\nonumber\\
&& \rule{0mm}{6mm}  + 2\at(2\au(1+\au+\at)-3\at)u^6 + ((9\at+\au\ad)(1+\au+\at)-4\at\au^2
-\au\ad^2-3\at\ad)u^5 \nonumber \\
&& \rule{0mm}{6mm}
+(4\ad(1+\au+\at)-16\at\au-2\ad^2-4\au^2\ad)u^4 + 
((1+\au+\at)\au-2\au^3-15\au\ad-12\at)u^3
\nonumber \\
&& \left.\left. \rule{0mm}{6mm}-(10\au^2+12\ad)u^2 -14\au u -6\right)\right]
\nonumber\\
H &=& \frac{r_+^4}{ L^5}\frac{1}{3 u^2 \HHH(u)}\left(\rule{0mm}{6mm}
(3(1+\au+\at)\at-\ad\at)u^5  + (4(1+\au+\at)\ad-8\at\au)u^4 \right.
\nonumber\\
&& \left. + (5(1+\au+\at)\au-7\au\ad-12\at)u^3 + (6 (1+\au+\at)-8\au^2-10\ad)u^2
-20\au u-12)\rule{0mm}{6mm}\right)
\nonumber\\
I &=& \frac{r_+^4}{L^5 u} \left[ 4(1-u)(1+(1+\au)u-\at u^2)\right]
\nonumber\\
J &=&\frac{r_+^4}{L^5 }\frac{1 }{u^2 \HHH(u)^{1/2}} 
\sqrt{(1-u)(1+(1+\au) u -\at u^2)\, }\, (3+2\au u + \ad u^2)
\eea

\end{document}